\documentclass[pra,showpacs,floatfix]{revtex4}
\usepackage{amsmath}
\usepackage{epsfig}
\usepackage{amssymb}
\usepackage{dcolumn}
\usepackage{graphicx}
\usepackage{dcolumn}
\usepackage{color}

\begin{document}

\title{Spontaneous symmetry breaking in Schr\"{o}dinger lattices with two
nonlinear sites}
\author{Valeriy A. Brazhnyi}
\email{brazhnyy@gmail.com}
\affiliation{Centro de F\'{\i}sica do Porto, Faculdade de Ci\^encias, Universidade do
Porto, R. Campo Alegre 687, Porto 4169-007, Portugal}
\author{Boris A. Malomed}
\email{malomed@post.tau.ac.il}
\affiliation{ICFO---Institut de Ciencies Fotoniques, Universitat Politecnica de
Catalunya, Mediterranean Technology Park, 08860 Castelldefels (Barcelona),
Spain\\
and Department of Physical Electronics, School of Electrical Engineering,
Faculty of Engineering, Tel Aviv University, Tel Aviv 69978, Israel\thanks{%
permanent address}}

\begin{abstract}
We introduce discrete systems in the form of straight (infinite) and
ring-shaped chains, with two symmetrically placed nonlinear sites.
The systems can be implemented in nonlinear optics (as waveguiding
arrays) and BEC (by means of an optical lattice). A full set of
exact analytical solutions for symmetric, asymmetric, and
antisymmetric localized modes is found, and their stability is
investigated in a numerical form. The symmetry-breaking bifurcation
(SBB), through which the asymmetric modes emerge from the symmetric
ones, is found to be of the subcritical type. It is transformed into
a supercritical bifurcation if the nonlinearity is localized in
relatively broad domains around two central sites, and also in the
ring of a small size, i.e., in effectively nonlocal settings. The
family of antisymmetric modes does not undergo bifurcations, and
features both stable and unstable portions. The evolution of
unstable localized modes is investigated by means of direct
simulations. In particular, unstable asymmetric states, which exist
in the case of the subcritical bifurcation, give rise to breathers
oscillating between the nonlinear sites, thus restoring an effective
dynamical symmetry between them.
\end{abstract}

\maketitle

\section{Introduction}

Spontaneous symmetry breaking is a fundamental effect caused by the
interplay of nonlinearity with linear potentials featuring basic symmetries,
such as double-well structures. In particular, while it is commonly known
that the ground state in one-dimensional quantum mechanics follows the
symmetry of the underlying double-well potential \cite{LL}, the
self-attractive nonlinearity added to the respective Schr\"{o}dinger
equation [which transforms it into the Gross-Pitaevskii equation for a
Bose-Einstein condensate (BEC)\ of interacting atoms loaded into the
double-well potential \cite{BEC}, or into the nonlinear Schr\"{o}dinger
equation in optical counterparts of the system \cite{NLS}] breaks the
symmetry of the ground state, replacing it by a new asymmetric state
minimizing the system's energy, provided that the strength of the
self-attraction exceeds a certain critical value (see, e.g., Ref. \cite{misc}
for the general consideration, and Ref. \cite{misc2} for the analysis of the
symmetry-breaking self-trapping in BEC). The spontaneous symmetry breaking
in double-well potentials was realized experimentally in BEC \cite{Markus}
and in nonlinear optics, where the symmetry breaking was observed in a setup
based on a photorefractive material \cite{photo}.

The \textit{symmetry-breaking bifurcation} (SBB), which destabilizes the
symmetric ground state and gives rise to an asymmetric one in the nonlinear
system, was originally discovered in a discrete model of self-trapping \cite%
{Chris}. In nonlinear optics, a similar SBB\ was predicted in Ref. \cite%
{Snyder} for continuous-wave (spatially uniform) states in the model of
dual-core fibers. For solitons in dual-core systems, this bifurcation was
studied in detail in Ref. \cite{dual-core}. Subsequently, the SBB was
studied for gap solitons in dual-core fiber Bragg gratings with the same
Kerr (cubic) nonlinearity as in the ordinary fibers \cite{Mak}. In another
physical setting which also features the cubic nonlinearity, the SBB\ was
predicted for matter-wave solitons in the self-attractive BEC loaded into a
dual-trough potential trap \cite{Arik}-\cite{Luca}.

The self-focusing cubic nonlinearity gives rise to the soliton bifurcations
of the \textit{subcritical} (alias \textit{backward}) type, in which the
branches of asymmetric modes emerge as unstable ones, going backward and
getting stabilized after switching their direction forward at turning points
\cite{bif}. On the other hand, the combination of the self-focusing
nonlinearity with a periodic potential acting in the free direction
(perpendicular to the direction of the action of the double-well potential)
changes the character of the bifurcation from subcritical to supercritical,
with the asymmetric branches emerging as stable ones and immediately going
in the forward direction \cite{Arik,Warsaw2}. The SBB in the model of the
dual-core fiber Bragg grating is of the forward type too \cite{Mak}. The
sub- and supercritical (alias backward/forward) SBBs may be regarded as
examples of phase transitions of the first and second kinds, respectively
\cite{bif}.

A physically interesting alternative to the linear double-well potential is
the setting with an effective \textit{pseudopotential} \cite{Harrison}
induced by the double-peak spatial modulation of the local nonlinearity
coefficient, which may be implemented in optics and BEC alike \cite%
{Pu,Barcelona}. The ultimate form of such a setting is the one with the
nonlinearity concentrated at two points, in the form of a symmetric pair of
delta-functions or narrow Gaussians \cite{Thawatchai,Nir}, as well as a
two-dimensional counterpart of the system, in the form of two parallel
troughs in which the nonlinearity is applied \cite{Hung} (the underlying
model, with the self-attractive nonlinearity represented by a single
delta-function, was introduced much earlier in Ref. \cite{Mark}). The SBB of
solitons in the symmetric double-well nonlinear pseudopotentials was studied
recently, featuring the subcritical type of the symmetry breaking \cite%
{Thawatchai,Hung,Nir}.

The spontaneous symmetry breaking was also analyzed for solitons in
dual-core discrete systems, with the uniform coupling between two parallel
chains \cite{Herring}, or with the coupling established at a single site
\cite{Ljupco}. In the former case, the SBB is subcritical, while in the
latter case it is supercritical.

The objective of the present work is to consider the SBB\ in one-dimensional
discrete lattices with the nonlinearity tightly concentrated at two
symmetric sites, or in narrow regions around them. This is a straightforward
discrete counterpart of the double-well nonlinear pseudopotential \cite%
{Thawatchai,Nir}, which offers a simple testbed for the study of
symmetry-breaking effects in discrete media. Physically, the linear chain
with two nonlinear sites can be readily implemented in optics, by embedding
two nonlinear cores into an arrayed linear waveguide, and in BEC, by means
of the Feshbach-resonance technique applied locally to the condensate
trapped in a deep optical lattice \cite{BEC}. An essential advantage of the
system is that, as we demonstrate below, it admits a fully analytical
solution (for the infinite chain), with an arbitrary separation between the
two symmetric nonlinear sites, while the stability of the exact solutions
may be efficiently predicted following general principles of the elementary
bifurcation theory \cite{bif} and the Vakhitov-Kolokolov (VK) criterion \cite%
{VK}.

The paper is organized as follows. The model is introduced in Section 2,
which is followed by producing exact solutions for symmetric, asymmetric,
and antisymmetric localized modes in Section 3. Numerical results, obtained
for finite lattices, are reported in Section 4 (in particular, it is
demonstrated that the SBB\ is subcritical in the system with two nonlinear
sites). In Section 5, we consider an essentially different version of the
system, in the form of a ring-shaped chain, with the two nonlinear sites
placed at diametrically opposite points, in which case an analytical
solution for the SBB is available too. The paper is concluded by Section 6.

\section{The model}

According to what is said above, the model is based on the linear discrete
Schr\"{o}dinger equation with two nonlinear sites embedded into it:%
\begin{equation}
i\frac{du_{n}}{dt}+(1/2)\left( u_{n+1}+u_{n-1}-2u_{n}\right) +\left( \delta
_{n,0}+\delta _{n,l}\right) \left\vert u_{n}\right\vert ^{2}u_{n}=0,
\label{1}
\end{equation}%
where $l$ is the integer distance between the two nonlinear sites, and $%
\delta _{n,m}$ is the Kronecker's symbol. The evolutional variable, $t$, is
time in the application to BEC, or propagation distance in the case of an
array of optical waveguides. In the former case, the nonlinearity at two
sites can be induced by focusing laser beams, that may enhance the
nonlinearity through the Feshbach resonance \cite{optical-FR}, at two
particular droplets of the condensate trapped in a deep optical lattice. In
the photonic realization of the setting, strong nonlinearity in two
particular cores in the waveguiding array can be readily imposed by doping
them with resonant atoms (see, e.g., Ref. \cite{doping}), which does not
affect the linear coupling of these sites to adjacent ones, as implied in
Eq. (\ref{1}). The same mechanisms can be used for inducing the local
nonlinearity in circular configurations considered below in Section 5.

Stationary solutions to Eq. (\ref{1}) are sought for as $u_{n}(t)=e^{-i%
\omega t}U_{n}$, where real stationary field $U_{n}$ obeys equation%
\begin{equation}
\omega U_{n}+(1/2)\left( U_{n+1}+U_{n-1}-2U_{n}\right) +\left( \delta
_{n,0}+\delta _{n,l}\right) U_{n}^{3}=0.  \label{U}
\end{equation}%
Along with this model, we will also consider its version with a smoothed
form of the discrete delta-function, \textit{viz}.,%
\begin{equation}
\delta _{n,n_{0}}\rightarrow \exp \left( -\left( n-n_{0}\right) ^{2}/\Delta
^{2}\right) ,  \label{Delta}
\end{equation}%
where $n_{0}=0$ or $l$, and $\Delta $ is the smoothing width.

A preliminary remark is that Eq. (\ref{U}) with $l=0$ corresponds to the
single nonlinear site with the double strength, placed at $n=0$. In that
case, an obvious exact solution is%
\begin{equation}
\left( U_{n}\right) _{\mathrm{single}}=Ae^{-\kappa |n|},~A^{2}=(1/2)\sinh
\kappa \text{,}  \label{single}
\end{equation}%
where $\kappa $ is connected to $\omega $ by the dispersion relation for
evanescent waves in the linear lattice:%
\begin{equation}
\omega =-2\sinh ^{2}\left( \kappa /2\right) .  \label{omega}
\end{equation}%
Taking relation (\ref{omega}) into regard, the norm of solution (\ref{single}%
) is%
\begin{equation}
N\equiv \sum_{n=-\infty }^{+\infty }U_{n}^{2}=(1/2)\cosh \kappa \equiv
\left( 1-\omega \right) /2.  \label{N}
\end{equation}%
Note that this expression for $N$ satisfies the VK stability criterion, $%
dN/d\omega <0$, hence solutions (\ref{single}) may be stable \cite{MIW}. On
the other hand, the continuous counterpart of Eq. (\ref{U}) with the single
nonlinear site is%
\begin{equation}
\omega U+\frac{1}{2}\frac{d^{2}U}{dx^{2}}+2\delta (x)U^{3}=0,  \label{cont0}
\end{equation}%
where $\delta (x)$ is the delta-function (this continuous equation was first
introduced in Ref. \cite{Mark}). It has an obvious localized solution,%
\begin{equation}
U(x)=\left( -\omega /2\right) ^{1/4}\exp \left( -\sqrt{-2\omega }|x|\right) ,
\label{deg}
\end{equation}%
whose norm is \emph{degenerate }(it does not depend on $\omega $): $N\equiv
\int_{-\infty }^{+\infty }U^{2}(x)dx=1/2$. Being formally neutrally stable
in terms of the VK criterion, all solutions (\ref{deg}) are actually \emph{%
unstable} \cite{Nir}. The degeneracy and instability of solutions (\ref{deg}%
) resemble the classical properties of the Townes' solitons \cite{Townes}.

\section{Exact solutions for the infinite lattice}

\subsection{General analysis}

Symmetric, antisymmetric, and asymmetric solutions to Eq. (\ref{U}) can be
found in an \emph{exact} form, following the pattern of the exact solutions
for the continuous counterpart of Eq. (\ref{U}), which was considered in
Ref. \cite{Thawatchai}:%
\begin{equation}
\omega U+\frac{1}{2}\frac{d^{2}U}{dx^{2}}+\left[ \delta \left( x\right)
+\delta \left( x-l\right) \right] U^{3}=0,  \label{cont}
\end{equation}%
cf. Eq. (\ref{cont0}). Exact symmetric and asymmetric solutions to the
discrete equation (\ref{U}) are sought in the following form:%
\begin{equation}
U_{n}=\left\{
\begin{array}{cc}
Ae^{\kappa n}, & \mathrm{at}~~n\leq 0, \\
B\cosh \left( \kappa \left( n-n_{0}\right) \right) , & \mathrm{at}~0\leq
n\leq l, \\
Ce^{-\kappa \left( n-l\right) }, & \mathrm{at}~n\geq l.%
\end{array}%
\right.  \label{ansatz}
\end{equation}%
Coordinate $n_{0}$ determines the location of the center of the intermediate
part of the solution. Note that $n_{0}$ does not need to be an integer
number. Symmetric and asymmetric modes correspond, respectively, to $%
n_{0}=l/2$ and $n_{0}\neq l/2$.

Ansatz (\ref{ansatz}) automatically satisfies the linear discrete Schr\"{o}%
dinger equation. There remains to check Eq. (\ref{U}) at the nonlinear sites
(of course, imposing the condition of the continuity of the solution at
these sites). With regard to Eq. (\ref{omega}), the continuity condition
yields relations between the amplitudes:%
\begin{equation}
B=\frac{A}{\cosh \left( \kappa n_{0}\right) }=\frac{C}{\cosh \left( \kappa
l-\kappa n_{0}\right) }~,  \label{ABC}
\end{equation}%
and the equation at the nonlinear sites amounts to the following relations:%
\begin{equation}
A^{2}=\frac{\sinh \kappa }{1+e^{-2\kappa n_{0}}},~C^{2}=\frac{\sinh \kappa }{%
1+e^{-2\kappa \left( l+n_{0}\right) }}.  \label{AC}
\end{equation}%
After some algebra, the condition that Eqs. (\ref{ABC}) and (\ref{AC}) yield
the same expression for $B$ produces an equation for $n_{0}$:%
\begin{gather}
x^{4}+\left( 3-L\right) x^{3}+\left( 1-3L^{-1}\right) x-L^{-2}=0,  \label{x}
\\
x\equiv e^{-2\kappa n_{0}},~L\equiv e^{2\kappa l}.  \label{L}
\end{gather}

\subsection{Symmetric modes}

\begin{figure}[h]
\epsfig{file=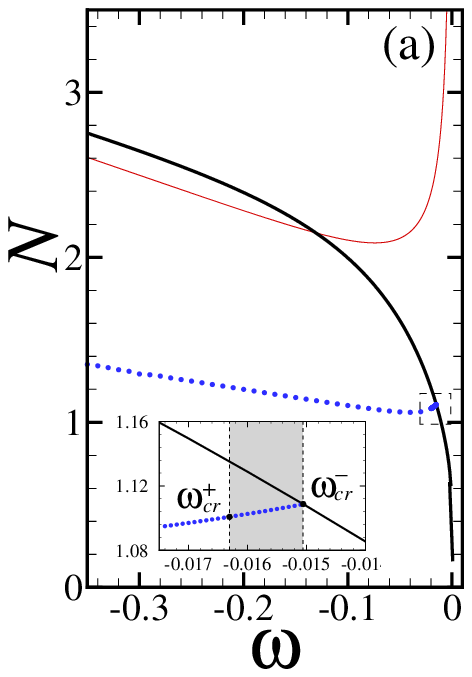,width=5cm}\epsfig{file=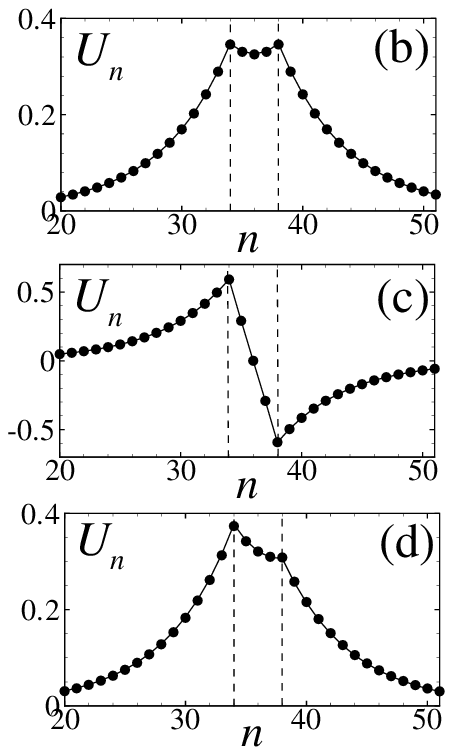,width=5cm}
\caption{(Color online) (a) The bifurcation diagram in the plane of the
intrinsic frequency of the localized modes ($\protect\omega $) and their
total norm ($N$), as produced by the analytical solution for the infinite
lattice with two symmetric nonlinear sites, separated by distance $l=4$. The
black thick, red thin and blue dotted lines correspond to the symmetric (%
\protect\ref{ABCsy}), antisymmetric (\protect\ref{ABCanti}) and asymmetric (%
\protect\ref{ABCas}) solutions, respectively. In the insert, the zoom of the
bifurcation of the branch of asymmetric solutions (the blue dotted line) is
shown. The gray area determines the region of the existence of the
two-peaked asymmetric solution, limited by critical frequencies $\protect%
\omega _{\mathrm{cr}}^{+}<\protect\omega <\protect\omega _{\mathrm{cr}}^{-}$%
. In panels (b)-(d), examples of analytically found symmetric, antisymmetric
and asymmetric modes at $\protect\omega =-0.016$ and $l=4$ are shown.}
\label{first}
\end{figure}

Two roots of quartic equation (\ref{x}) are $x=\pm L^{-1/2}$. The negative
one is unphysical, while the positive root corresponds, according to Eq. (%
\ref{L}), to $n_{0}=l/2$, which represents the symmetric mode. The
amplitudes of the symmetric solution, as given by Eqs. (\ref{ABC}) and (\ref%
{AC}), are
\begin{subequations}
\label{ABCsy}
\begin{eqnarray}
A^{2} &=&C^{2}=\frac{\sinh \kappa }{1+e^{-\kappa l}}, \\
B^{2} &=&\frac{4e^{-\kappa l}\sinh \kappa }{\left( 1+e^{-\kappa l}\right)
^{3}}~.
\end{eqnarray}%
It is relevant to note that, in the case of $l=0$, Eqs. (\ref{ABCsy}) yield $%
A^{2}=B^{2}=C^{2}=(1/2)\sinh \kappa $, which naturally coincides with $A^{2}$
as given by Eq. (\ref{single}).

\subsection{Asymmetric modes}

Two other roots of Eq. (\ref{x}) represent a pair of \textit{asymmetric modes%
}:
\end{subequations}
\begin{equation}
x_{\pm }=\frac{1}{2}\left[ \left( L-3\right) \pm \left( L-1\right) \sqrt{%
\frac{L-4}{L}}\right] .  \label{as}
\end{equation}%
Taking into regard Eq. (\ref{L}), it is easy to check that roots (\ref{as})
correspond to two values $\left( n_{0}\right) _{\pm }$ which are located
mutually symmetrically around the center: $x_{+}x_{-}=L^{-1}$, i.e., $\left(
n_{0}\right) _{+}+\left( n_{0}\right) _{-}=l$. Further, roots (\ref{as}) are
physical if they are real and positive, which means $L>4$. Thus, the \textit{%
symmetry-breaking bifurcation} (SBB), i.e., the appearance of the asymmetric
modes from the symmetric one [see Fig. \ref{first}(a)], happens, with the
increase of distance $l$ between the nonlinear sites (at fixed $\kappa $,
i.e., fixed $\omega $), at the \textit{critical point}, $\kappa _{\mathrm{cr}%
}l=\ln 2$, or, in terms of the frequency of the localized mode, at%
\begin{equation}
\omega =\omega _{\mathrm{cr}}^{-}=-2\sinh ^{2}\left( \frac{\ln 2}{2l}\right)
.  \label{cr}
\end{equation}%
In other words, for fixed $l$, the SBB occurs with the increase of $%
\left\vert \omega \right\vert $ at point (\ref{cr}), the asymmetric modes
existing at $\left\vert \omega \right\vert >\left\vert \omega _{\mathrm{cr}%
}^{-}\right\vert $.

The asymmetric mode keeps the double-peak shape if the minimum point in
expression (\ref{ansatz}), $n_{0}$, is a really existing minimum, rather
than a virtual one, i.e., it stays in interval $0<n_{0}<l$. Further, this
means that $x_{\pm }$ must fall into the following interval: $L^{-1}<x_{\pm
}<1$. It is easy to check, making use of Eq. (\ref{as}), that the latter
condition places $L$ and $l$ into narrow intervals of their values, namely, $%
4<L<2+\sqrt{5}\approx 4.24$, and, accordingly,%
\begin{equation}
\ln 2\approx 0.69<\kappa l<(1/2)\ln \left( 2+\sqrt{5}\right) \approx 0.72,
\label{l}
\end{equation}%
cf. Eq. (\ref{cr}). At $\kappa l>\left( 1/2\right) \ln \left( 2+\sqrt{5}%
\right) $, i.e., $\omega <\omega _{\mathrm{cr}}^{+}=-2\sinh ^{2}\left[
\left( 4l\right) ^{-1}\ln \left( 2+\sqrt{5}\right) \right] $ [in other
words, at $|\omega |>\left\vert \omega _{\mathrm{cr}}^{+}\right\vert $, cf.
Eq. (\ref{cr})], $n_{0}$ leaves the region of $0<n_{0}<l$, hence the
asymmetric mode becomes single-peaked (``strongly asymmetric"). The
transition from the double-peak asymmetric mode to the single-peak shape is
illustrated by Fig. \ref{fig:asym}(b).

The final expressions for the amplitudes of the asymmetric mode are obtained
by the substitution of expressions (\ref{as}) and (\ref{L}) into Eqs. (\ref%
{ABC}) and (\ref{AC}):
\begin{subequations}
\label{ABCas}
\begin{gather}
\left\{ A^{2},C^{2}\right\} =\left\{ \frac{2e^{\kappa l}\sinh \kappa }{%
\left( e^{2\kappa l}-1\right) \left( e^{\kappa l}\pm \sqrt{e^{2\kappa l}-4}%
\right) }\right\} \\
B^{2}=\frac{16e^{2\kappa l}\left[ e^{\kappa l}\left( e^{2\kappa l}-3\right)
\pm \left( e^{2\kappa l}-1\right) \sqrt{e^{2\kappa l}-4}\right] \sinh \kappa
}{\left( e^{2\kappa l}-1\right) ^{3}\left( e^{\kappa l}\pm \sqrt{e^{2\kappa
l}-4}\right) ^{3}}.
\end{gather}%
where signs $\pm $ pertain to the two conjugate asymmetric modes. Note that,
at the SBB point, $e^{\kappa l}=2$, expressions (\ref{ABCas}) coincide with
their counterparts (\ref{ABCsy}) obtained above for the symmetric mode: at
this point, the amplitudes are $A^{2}=C^{2}=\left( 2/3\right) \sinh \kappa $
and $B^{2}=\left( 16/27\right) \sinh \kappa $.

As a natural measure of their asymmetry, we will use
\end{subequations}
\begin{equation}
\Theta \equiv \frac{A^{2}-C^{2}}{A^{2}+C^{2}}=\pm \sqrt{1-4e^{-2\kappa l}}.
\label{Theta}
\end{equation}%
Obviously, $\Theta =0$ at the SBB point, $e^{-\kappa l}=1/2$, and expression
(\ref{Theta}) makes sense past the bifurcation point, i.e., at $e^{-\kappa
l}<1/2$. In particular, at the above-mentioned point of the transition from
the double- to single-peak profile, $\kappa l=\left( 1/2\right) \ln \left( 2+%
\sqrt{5}\right) $, the asymmetry is still relatively small, $\left\vert
\Theta \right\vert =\sqrt{5}-2\approx \allowbreak 0.24$.

\subsection{Antisymmetric modes}

Antisymmetric solutions, and their counterparts with broken antisymmetry (if
any) are looked for by dint of the following ansatz, cf. Eq. (\ref{ansatz}):
\begin{equation}
U_{n}=\left\{
\begin{array}{cc}
Ae^{\kappa n}, & \mathrm{at}~~n\leq 0, \\
B\sinh \left( \kappa \left( n-n_{0}\right) \right) , & \mathrm{at}~0\leq
n\leq l, \\
U_{n}=Ce^{-\kappa \left( n-l\right) }, & \mathrm{at}~n\geq l.%
\end{array}%
\right.  \label{ansatz2}
\end{equation}%
Looking for a bifurcation occurring to the antisymmetric mode, one arrives
at the respective roots for $x$ [with $x$ defined as in Eq. (\ref{L})]:%
\begin{equation}
x_{\pm }=\frac{1}{2}\left[ -\left( L-3\right) \pm \left( L-1\right) \sqrt{%
\frac{L-4}{L}}\right] ,  \label{unphys}
\end{equation}%
which differs from Eq. (\ref{as}) by the opposite sign in front of the first
term in the square brackets. This difference makes both roots (\ref{unphys})
negative (unphysical), hence the antisymmetric solutions \emph{do not}
undergo the bifurcation, similar to the situation known in the continuous
model (\cite{Thawatchai}).

The mode with the unbroken antisymmetry is given by the exact solution in
the form of ansatz (\ref{ansatz2}) with amplitudes
\begin{subequations}
\label{ABCanti}
\begin{gather}
A=-C=\sqrt{\frac{\sinh \kappa }{1-e^{-\kappa l}}}, \\
B=-\frac{2\sqrt{\sinh \kappa }e^{-\kappa l/2}}{\left( 1-e^{-\kappa l}\right)
^{3/2}}.
\end{gather}%
Finally, it is natural to call the antisymmetric modes, corresponding to
even and odd $l$, \textit{on-site} and \textit{inter-site} ones,
respectively, as they have the zero point, $n_{0}=l/2$, either coinciding or
not with the (central) site of the lattice.

\section{Numerical results for a finite lattice}

The objective of the numerical solution is to obtain solutions of stationary
equation (\ref{U}) for a finite lattice, compare them to the exact solutions
found above for the infinite lattice and identify stability of the
symmetric, asymmetric and antisymmetric solutions. In addition, the
numerical calculations help to identify the type of the SBB\ (sub- or
supercritical), as the analytical solution produces a very cumbersome
result, in this respect. Numerical stationary solutions were constructed in
the lattice of 71 sites. Highly asymmetric modes were obtained by means of
the continuation in $\omega $, using the Newton's iteration procedure that
started from the analytical asymmetric solution given by Eqs. (\ref{ansatz})
and (\ref{ABCas}). For all numerical solutions, we analyzed the linear
stability by solving the corresponding linear eigenvalue problem. The
results were checked by direct simulations of the underlying equation (\ref%
{1}).

\subsection{Symmetric modes}

In Fig. \ref{fig:sym}, families of numerically generated symmetric solutions
are shown for different distances ($l$) between the two nonlinear sites. As
said above, the stability of the numerically found solutions was identified
through the calculation of the corresponding eigenvalues of small
perturbations. The borders between stable and unstable segments of the
solution branches correspond to the SBB which destabilizes the symmetric
solutions. The numerical solutions are practically indistinguishable from
their analytical counterparts (\ref{ansatz}), (\ref{ABCsy}), therefore the
analytically found curves are not plotted separately (strictly speaking, the
numerical results cannot be identical to the analytical ones, as the
numerical computations were performed for the finite chain, while the
analytical findings pertain to the infinite one).

\begin{figure}[th]
\epsfig{file=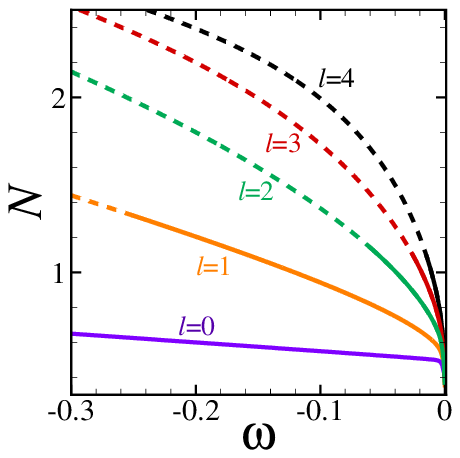,width=6cm}\epsfig{file=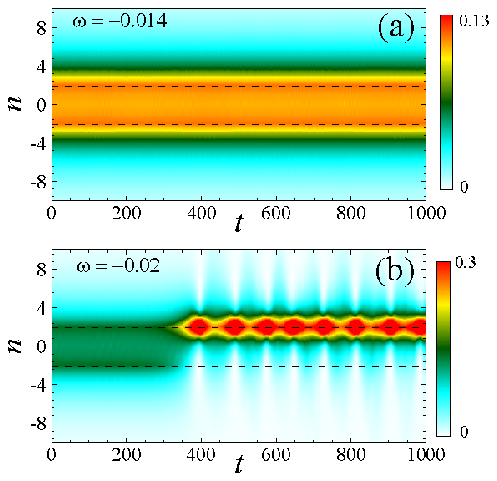,width=6cm}
\caption{(Color online) \textit{Left panel}: Families of numerically
generated symmetric solutions (their analytical counterparts are
indistinguishable from the numerical ones). The corresponding values
of the distance between the nonlinear sites, $l$, are indicated. The
solid and dashed lines depict, respectively, stable and unstable
portions of the solution families. \textit{Right panel}: The
evolution of stable (a) and unstable (b) symmetric modes with $l=4$,
taken at $\protect\omega =-0.014$ and $\protect\omega =-0.02$,
respectively. Here and in similar plots below, the spatiotemporal
evolution is displayed by means of the density contour plots. Small
perturbations ($\sim 1\%$) were added to initiate the evolution. Dashed lines show the position of nonlinear sites.}
\label{fig:sym}
\end{figure}

To check the predictions of the linear stability analysis, we simulated the
underlying equation (\ref{1}), with initial profiles for symmetric modes
taken in regions where these solutions are expected to have different
stability. Typical examples, presented in right panel of Fig. \ref{fig:sym}, show that the
unstable symmetric stationary solution transforms into a pulsating
single-peak mode, which breaks the symmetry, getting spontaneously localized
on one of the nonlinear sites.

\subsection{Asymmetric modes}

The numerical results for the asymmetric modes are displayed in Fig. \ref%
{fig:asym}(a). In particular, unstable parts of the families of asymmetric
solutions are those which are related to the SBB of the \emph{subcritical}
type [see Fig. \ref{asymm_theta}(b) below]. The SBB displayed by the
numerical results closely follows the analytical solution for the infinite
lattice. In particular, generic numerically found profiles of the
double-peaked (above $\omega _{\mathrm{cr}}^{+}$) and single-peak (below $%
\omega _{\mathrm{cr}}^{+}$) asymmetric modes are compared with their
analytical counterparts in Fig. \ref{fig:asym}(b), the difference between
the numerical and analytical solutions being $\lesssim 1\%$.

\begin{figure}[th]
\epsfig{file=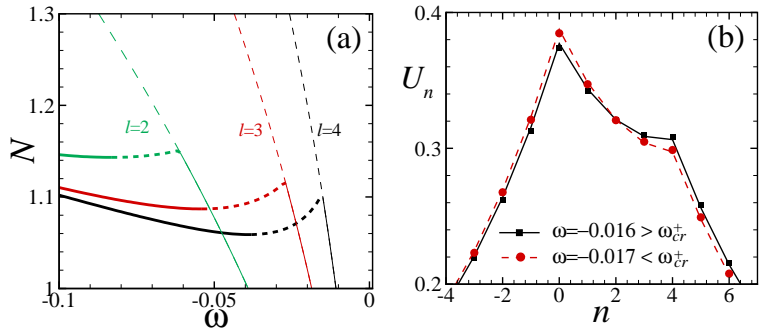,width=10cm}
\caption{(Color online) (a): The bold and thin lines show, respectively, the
families of asymmetric and symmetric solutions. As in Fig. \protect\ref%
{fig:sym}, solid and dashed lines corresponds to stable and unstable
portions of the solution families. In fact, the lines for the symmetric
solutions are identical to those displayed in Fig. \protect\ref{fig:sym}.
(b): An example of the transition from the double-peak shape for $\protect%
\omega >\protect\omega _{cr}^{+}$ to the single-peak one for $\protect\omega %
<\protect\omega _{cr}^{+}$, obtained from numerical (circles and
squares) and analytical (red dashed and black solid lines)
solutions.} \label{fig:asym}
\end{figure}

In Fig. \ref{asymm_theta} the numerical results obtained for the asymmetric
solutions are collected in the form of plots showing the asymmetry measure
defined, as per Eq. (\ref{Theta}), through the difference between squared
amplitudes of the solution at the two nonlinear sites, as a function of
frequency $\omega $ and total norm $N$. The corresponding discrepancy
between the numerical and analytical results is $<1\%$. In panel (a), the
numerically found bifurcation points are compared to those predicted
analytically by Eq. (\ref{cr}) (they correspond to values of $\omega _{%
\mathrm{cr}}^{-}$ indicated in the panel), which demonstrates a precise
agreement. An important conclusion clearly suggested by Fig. \ref%
{asymm_theta}(b) is that the bifurcation has the \textit{subcritical}
character, with the asymmetric branches being unstable exactly between the
SBB and turning points, as might be expected.

\begin{figure}[th]
\epsfig{file=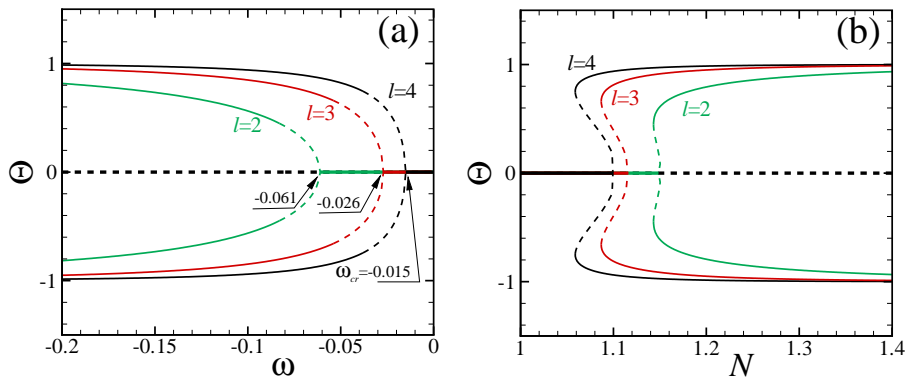,width=12cm} 
\caption{(Color online)
The family of the numerically found asymmetric solutions in the
planes of $(\Theta ,\protect\omega )$ (a) and $(\Theta ,N)$ (b) for
different distances between the nonlinear sites ($l$). As above, the
stable and unstable portions of the solution families are depicted
by solid and dashed lines, respectively.}
\label{asymm_theta}
\end{figure}

To check the predictions of the linear-stability analysis, we
simulated the underlying equation (\ref{1}) with initial profiles
for asymmetric modes taken in the regions where these solutions are
expected to be stable and unstable, respectively. Typical examples,
presented in Fig. \ref{fig1}, show that the unstable asymmetric
stationary solution transforms itself, after a transient period,
into a robust breather oscillating between two asymmetric
configurations (in that sense, the breathers restores an effective \textit{%
dynamical symmetry}).
%, while the unstable symmetric solution transforms into
%a pulsating single-peak mode, which gets spontaneously localized on one of
%the nonlinear sites.

\begin{figure}[th]
\epsfig{file=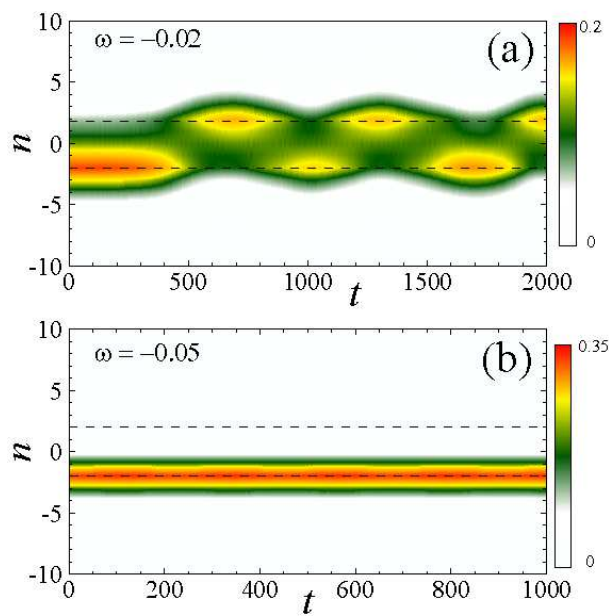,width=8cm}
\caption{((Color online) The evolution of unstable (a) and stable (b)
asymmetric modes with $l=4$, taken at $\protect\omega =-0.02$ and $\protect%
\omega =-0.05$, respectively. Dashed lines show the position of nonlinear sites.} 
\label{fig1}
\end{figure}

\subsection{Antisymmetric modes}

Families of numerically found on-site (a) and inter-site (b) antisymmetric
modes are shown in Fig. \ref{fig:antisym}. As in the previous cases the
curves, they are practically indistinguishable from the analytically found
counterparts. In contrast to the symmetric and asymmetric solutions, norm $N$
of the antisymmetric ones is bounded by a minimum value (the existence
threshold). It is also seen that each curve features stable and unstable
portions, the border between which approaches the bottom of the curve with
the increase the distance between the nonlinear sites, $l$. Panels (c) and
(d) in Fig. \ref{fig:antisym} display typical profiles of the on-site and
inter-site antisymmetric modes, again showing very good agreement with the
respective analytically predicted profiles. Note that, according to panels
(a) and (b), the antisymmetric solution shown for $l=4$ is stable, while the
ones for $l=1,2,3$ are unstable.

\begin{figure}[th]
\epsfig{file=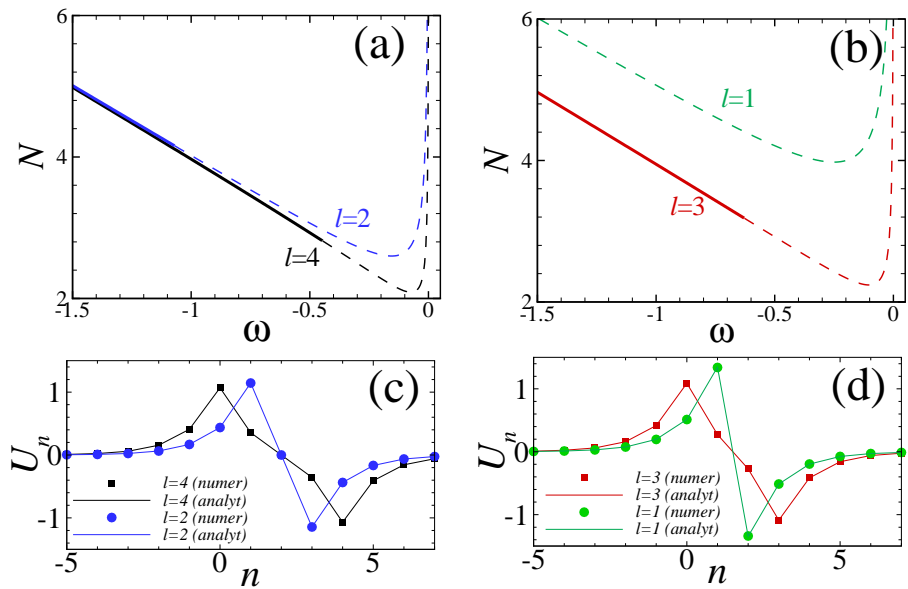,width=12cm} 
\caption{(Color
online) Panels (a) and (b) display $N(\protect\omega )$ curves for
families of the antisymmetric solutions of on-site and inter-site
types, respectively (i.e., even or odd distance $l$ between the
nonlinear sites). The solid and dashed portions of the curves refer,
as usual, to stable and unstable solutions. Panels (c) and (d)
display typical profiles of the antisymmetric on-site and inter-site
modes, as obtained in the numerical and analytical forms at
$\protect\omega =-0.5$.}
\label{fig:antisym}
\end{figure}

To check the linear-stability predictions for the antisymmetric modes, we
took initial profiles of the antisymmetric solutions from Fig. \ref%
{fig:antisym}(c),(d) at $\omega =-0.5$, when the on-site/inter-site modes
with $l=1,2,3$ are unstable, while the one with $l=4$ is stable, and
simulated Eq. (\ref{1}) with small initial perturbations ($\sim 1\%$). The
results are presented in Fig. \ref{antisym_dyn}. The unstable solutions
spontaneously transform into single-peak modes localized on one of the
nonlinear sites, which demonstrates decaying oscillations of the amplitude.
On the other hand, the linearly stable antisymmetric mode, with $l=4$, is
indeed robust in the direct simulations.

\begin{figure}[th]
\epsfig{file=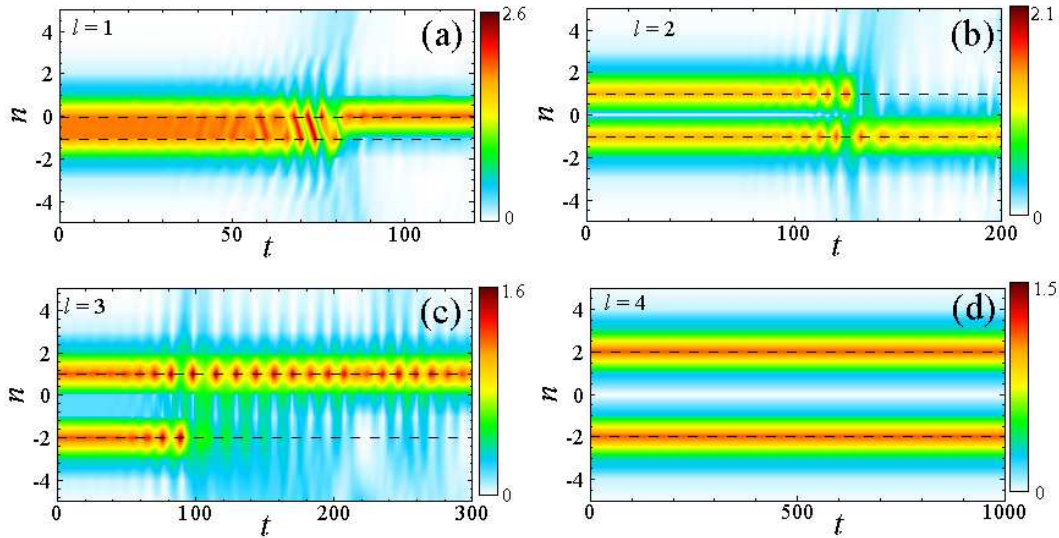,width=14cm}
\caption{(Color
online) The evolution of unstable (a), (b), (c) and stable (d)
antisymmetric modes. Initial profiles are taken from Fig.
\ref{fig:antisym}(c),(d) with $\protect\omega =-0.5$, for
$l=1,2,3,4$.} \label{antisym_dyn}
\end{figure}

\subsection{Effects of the finite extension of the nonlinear region}

In the model with two finite nonlinearity domains, defined as per Eq. (\ref%
{Delta}), results were obtained in the numerical form. They are displayed
for the families of symmetric and asymmetric modes in Fig. \ref{gauss_sym},
and for the antisymmetric ones in Fig. \ref{gauss_antisym}. In particular,
Fig. \ref{gauss_sym} demonstrates that the increase of the width of the
nonlinearity domain transforms the subcritical bifurcation into a \textit{%
supercritical} one, thus completely stabilizing the asymmetric modes and
making the value of $N$ at the SBB\ point lower. It is also worthy to note
that, as seen in Fig. \ref{gauss_antisym}, the broader nonlinearity gives
rise to an additional stability window close to the existence threshold
(minimum value of $N$).

\begin{figure}[th]
\epsfig{file=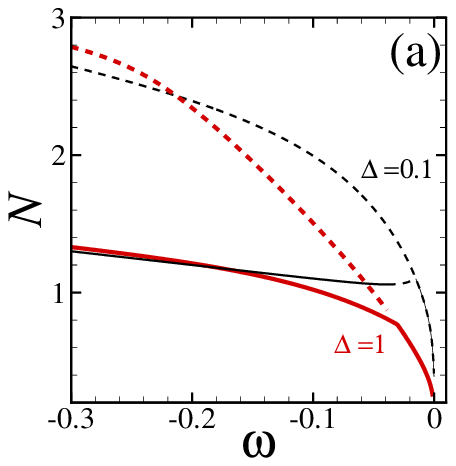,height=4.5cm}\epsfig{file=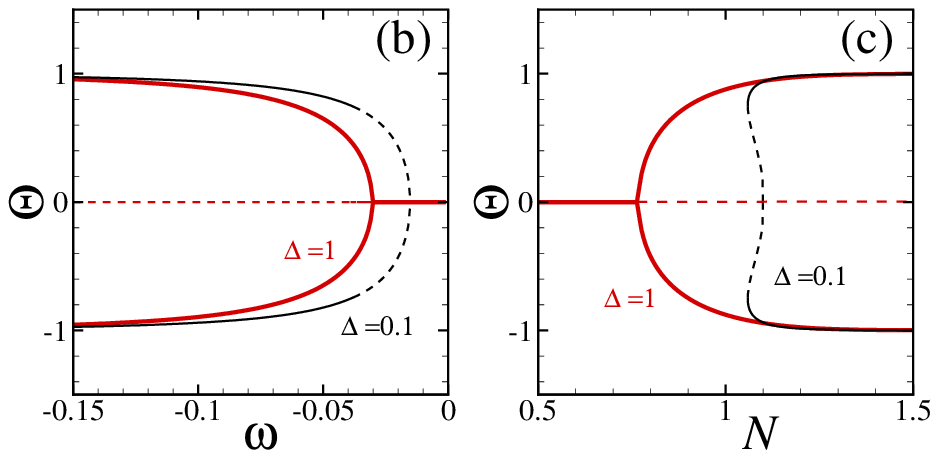,height=4.5cm}
\caption{(Color online) (a) Curves $N(\protect\omega )$ of the symmetric and
asymmetric solution families for $l=4$ with $\Delta =0.1$ in Eq. (\protect
\ref{Delta}) (black thin lines, which are actually tantamount to their
counterparts with the $\protect\delta $-like nonlinearity), and with $\Delta
=1$ (red thick lines), which correspond to a relatively broad Gaussian. In
(b) and (c), the corresponding asymmetry ratio $\Theta $ is shown versus $%
\protect\omega $ and $N$. }
\label{gauss_sym}
\end{figure}

\begin{figure}[th]
\epsfig{file=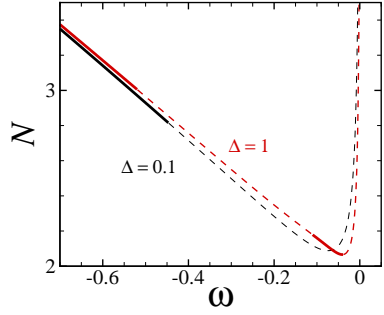,width=5cm}
\caption{(Color
online) The $N(\protect\omega )$ curves of antisymmetric solution
families for $l=4$ with $\Delta =0.1$ (black curves) and $\Delta =1$
(red curves), cf. Fig. \protect\ref{gauss_sym}.}
\label{gauss_antisym}
\end{figure}

\section{The ring model}

If the infinite linear chain is replaced by a circle (ring) with the
nonlinear sites placed at diametrically opposite points, Eq. (\ref{U}) is
replaced by the following one:
\end{subequations}
\begin{equation}
\omega U_{n}+\frac{1}{2}\left( U_{n+1}+U_{n-1}-2U_{n}\right) +\left( \delta
_{n,n_{-}}+\delta _{n,n_{+}}\right) U_{n}^{3}=0,  \label{diameter}
\end{equation}%
where $n=n_{\pm }$ may be defined as the top and bottom points of a vertical
diameter cutting the circle. In fact, it is more convenient to replace Eq. (%
\ref{diameter}) by a system of two identical equations for two semi-circles,
left and right ones. In each equation, $n$ takes values $n_{-}<n<n_{+}$ ($%
n_{-}<0$ and $n_{+}>0$ are assumed). %\begin{figure}[th]
%\epsfig{file=ring_f.eps,width=10cm} \caption{(Color online) Examples
%of ring configurations.} \label{ring}
%\end{figure}
In BEC, the circular chain can be realized as a combination of a toroidal
trap and periodic potential created in it \cite{torus}, and in optics it
corresponds to an array of waveguides created in a hollow cylindrical shell,
or written in the form of the ring in a bulk sample \cite{Jena}.

In principle, one may expect two types of the symmetry breaking in this
setting: between the top and bottom, which is a counterpart of what was
considered above in the framework of the rectilinear chain, or between the
left and right semi-circles.

The linear solutions for the left ($l$) and right ($r$) semi-circles can be
looked for as%
\begin{equation}
\left( U_{n}\right) _{l,r}=A_{l,r}\cosh \left( \kappa \left(
n-n_{l,r}\right) \right) ,  \label{lr}
\end{equation}%
with $\kappa $ related to $\omega $ by Eq. (\ref{omega}), and some constants
$A_{l,r}$ and $n_{l,r}$ ($n_{l}$ and $n_{r}$ are not necessarily integer
numbers). Then, the continuity conditions should be imposed at points $%
n=n_{\pm }$, where the two semi-circles are linked into the entire circle:%
\begin{equation}
A_{l}\cosh \left( n_{\pm }-n_{l}\right) =A_{r}\cosh \left( n_{\pm
}-n_{r}\right) .  \label{link}
\end{equation}%
After simple manipulations, one may eliminate the amplitudes from two
equations (\ref{link}), which leads to the following consistency condition: $%
\cosh \left( \kappa \left( n_{+}-n_{-}-n_{l}+n_{r}\right) \right) =\cosh
\left( \kappa \left( n_{+}-n_{-}+n_{l}-n_{r}\right) \right) .$ It is obvious
that the consistency condition can be met in the case of $n_{+}=n_{-}$,
which is trivial (zero length of the ring), or
\begin{equation}
n_{l}=n_{r}\equiv n_{0}.  \label{0}
\end{equation}%
Further, it then follows from Eqs. (\ref{link}) that a consequence of Eq. (%
\ref{0}) is $A_{l}=A_{r}\equiv A$, hence the symmetry breaking between the
left and right circles is impossible.

However, the top-bottom symmetry breaking is possible. Substituting ansatz (%
\ref{lr}) with the left-right symmetry into Eq. (\ref{diameter}) at points $%
n=n_{\pm }$, the result of a straightforward analysis is a system of two
equations, corresponding to $+$ and $-$:%
\begin{equation}
\tanh \left( \kappa \left( n_{\pm }-n_{0}\right) \right) \left[ 1-\tanh
^{2}\left( \kappa \left( n_{\pm }-n_{0}\right) \right) \right] =\pm
A^{2}/\sinh \kappa \text{,}  \label{tanh}
\end{equation}%
where $n_{0}$ is defined as per Eq. (\ref{0}). In equations (\ref{tanh}), $%
A^{2}$ and $n_{0}$ are considered as two unknowns, for given $\kappa $ and $%
n_{\pm }$. As follows from Eqs. (\ref{tanh}), the symmetric solution, with $%
n_{0}=\left( 1/2\right) \left( n_{+}+n_{-}\right) $, has amplitude%
\begin{equation}
A_{\mathrm{symm}}^{2}=\left( \sinh \kappa \right) \tanh \left( \left( \kappa
/2\right) \left( n_{+}-n_{-}\right) \right) \left[ 1-\tanh ^{2}\left( \left(
\kappa /2\right) \left( n_{\pm }-n_{0}\right) \right) \right] .
\label{Asymm}
\end{equation}

An obvious corollary of Eqs. (\ref{tanh}) is relation%
\begin{equation}
\tanh \left( \kappa \left( n_{+}-n_{0}\right) \right) \left[ 1-\tanh
^{2}\left( \kappa \left( n_{+}-n_{0}\right) \right) \right] -\tanh \left(
\kappa \left( n_{0}-n_{-}\right) \right) \left[ 1-\tanh ^{2}\left( \kappa
\left( n_{0}-n_{-}\right) \right) \right] =0.  \label{tanhtanh}
\end{equation}%
The SBB point can be found by setting $n_{0}=\left( 1/2\right) \left(
n_{+}-n_{-}\right) +\delta n_{0}$, with infinitesimal $\delta n_{0}$, and
demanding that the coefficient in front of $\delta n_{0}$ in the respective
expansion of the left-hand side of Eq. (\ref{tanhtanh}) vanishes. The result
is that this happens at point $\cosh \left( \left( \kappa /2\right) \left(
n_{+}-n_{-}\right) \right) =\sqrt{3/2},$ which is equivalent to
\begin{equation}
\left[ \kappa \left( n_{+}-n_{-}\right) \right] _{\mathrm{cr}}=\ln \left( 2+%
\sqrt{3}\right) .  \label{SBB}
\end{equation}%
At this point, the amplitude is%
\begin{equation}
A_{\mathrm{cr}}^{2}=\left( 2/3^{3/2}\right) \sinh \kappa .  \label{Acr}
\end{equation}%
The asymmetric state exists, for given $\kappa $, i.e., given $\omega $, for
the length of the semi-circle, $n_{+}-n_{-}$, which is larger than the one
corresponding to Eq. (\ref{SBB}). Alternatively, at given $n_{+}-n_{-}$, the
asymmetric state exists for $\kappa $ exceeding the value defined by Eq. (%
\ref{SBB}).

\begin{figure}[th]
\epsfig{file=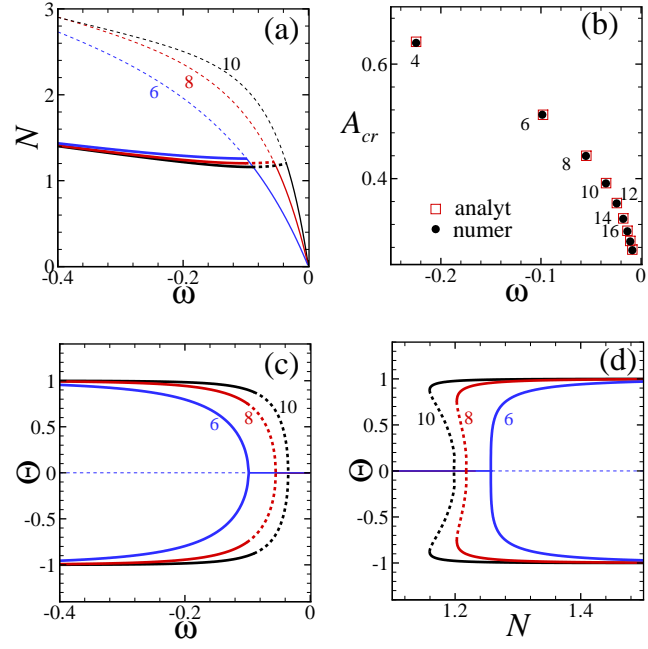,width=9cm} 
\caption{(Color online) (a): The
family of symmetric (thin) and asymmetric (thick) solutions for
three different lengths of the ring (6, 8, and 10). The solid and
dashed lines correspond to the stable and unstable solutions,
respectively. (b): The amplitude of the symmetric solution at the
bifurcation point, for different lengths of the ring (indicated by
numbers at the points). (c), (d): The dependence of asymmetry
$\Theta $ on the frequency (c) and norm $N$ (d), showing the change
of the bifurcation type in the ring with the increase of the ring's
size.} \label{ring}
\end{figure}

In Fig. \ref{ring} we summarize the results related to the ring
configuration. In panel (a), families of the symmetric and asymmetric
solutions are displayed for three different lengths of the ring. We have
checked the linear stability of the corresponding solutions through the
numerical computation of the eigenvalues for small perturbations. The
respective stable and unstable regions of the existence curves are shown by
the solid and dashed lines, respectively. As in the case of the linear
chain, at critical frequency $\omega _{\mathrm{cr}}$ calculated from Eq. (%
\ref{SBB}), the branches of the asymmetric solutions bifurcate from the
branch of symmetric solutions. The amplitude of the symmetric solutions at
the bifurcation point is displayed in Fig. \ref{ring}(b) for different
lengths of the ring.
%In Fig. \ref{ring}(b), we compare the amplitudes of the
%symmetric solutions, for different lengths of the ring, at the bifurcation
%point with ones calculated numerically and found very good agreement.

Also for different lengths of the ring, we have analyzed the type of the
corresponding SBB, calculating the asymmetry parameter $\Theta $ as per Eq. (%
\ref{Theta}). The dependences of $\Theta $ on frequency $\omega $ and norm $%
N $ are shown in Figs. \ref{ring}(c) and (d). As one can see, the length of
the ring plays a crucial role in the determination of the bifurcation type.
In the small ring (e.g., of length 6), the bifurcation is supercritical,
while the increase the length to $\geq 8$ changes it into a subcritical one.

\begin{figure}[th]
\epsfig{file=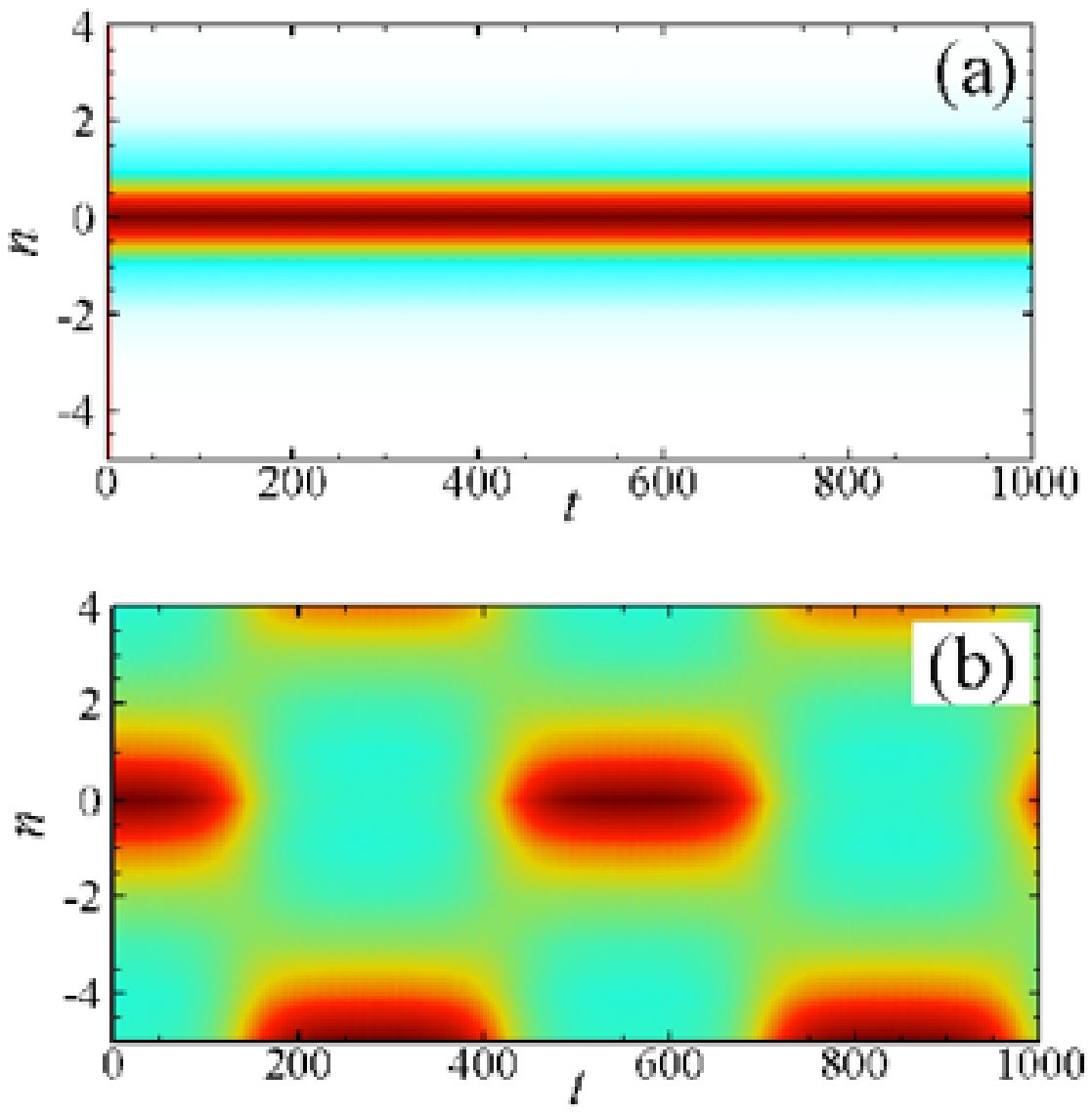,width=8cm}
\caption{(Color online) The evolution of stable (a) and unstable (b)
asymmetric solutions with initial profiles taken at $\protect\omega =-0.5$
and $\protect\omega =-0.05$, respectively. The length of the ring is 10
sites. Initial profiles were taken with small perturbations.}
\label{dyn_ring_asym}
\end{figure}

\begin{figure}[th]
\epsfig{file=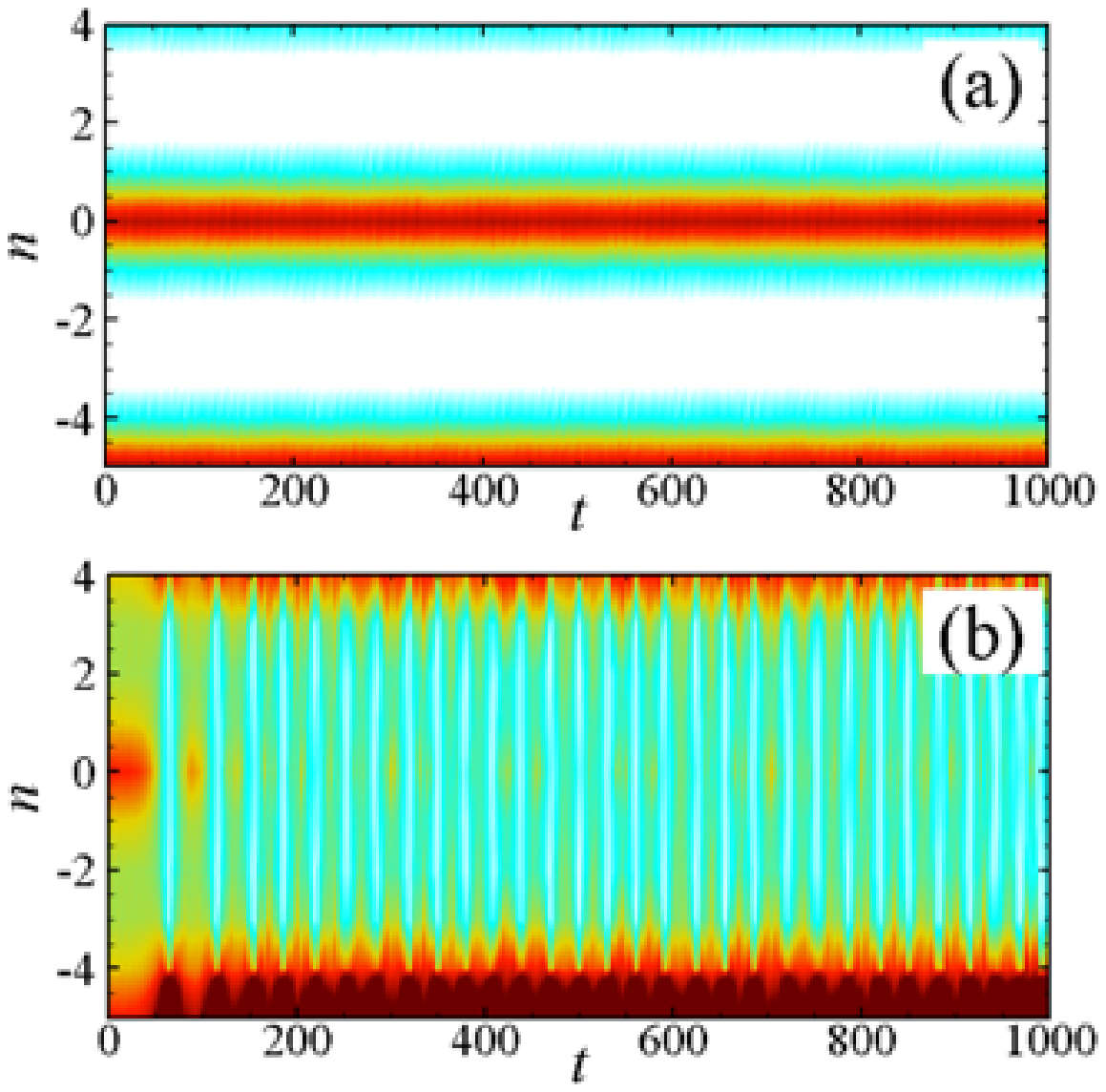,width=8cm}
\caption{(Color online) The evolution of stable (a) and unstable (b)
symmetric solutions with initial profiles taken at $\protect\omega =-0.05$
and $\protect\omega =-0.03$, respectively. The length of the ring is 10
sites. Initial profiles were taken with small perturbations.}
\label{dyn_ring_sym}
\end{figure}

In Figs. \ref{dyn_ring_asym} and \ref{dyn_ring_sym}, the evolution of stable
and unstable asymmetric and symmetric solutions is shown. The direct
simulations corroborate the linear-stability analysis. As one can see, the
solutions taken on stable parts of the existence curves are stable indeed.
The unstable asymmetric solution [see Fig. \ref{dyn_ring_asym}(b)] starts to
oscillate between two asymmetric configurations, while the unstable
symmetric solution in Fig. \ref{dyn_ring_sym}(b) rapidly transforms into an
oscillating single-peak mode. It is relevant to stress that, similar to the
situation in the straight chain (cf. Fig. \ref{fig1}), the instability of
symmetric modes transforms them into asymmetric ones, while the instability
of asymmetric modes leads to the emergence of the effective dynamical
symmetry.

Finally, it is relevant to note that, in the case of the circle of a finite
length, it is also possible to consider values $\omega >0$, i.e., imaginary
wavenumbers, $\kappa =i|\kappa |$, see Eq. (\ref{omega}). This means that
the exponentially decaying discrete waves are replaced by oscillatory ones.
After a simple analysis, the respective version of Eq. (\ref{tanh}) can be
derived in the following form:%
\begin{equation}
\tan \left( |\kappa |\left( n_{\pm }-n_{0}\right) \right) \left[ 1+\tan
^{2}\left( |\kappa |\left( n_{\pm }-n_{0}\right) \right) \right] =\mp A^{2}/%
\left[ 2\sin \left( |\kappa |\right) \right] \text{.}  \label{tan}
\end{equation}%
A straightforward analysis of Eq. (\ref{tan}) demonstrates that, on the
contrary to Eq. (\ref{tanh}), it \emph{does not} give rise to the symmetry
breaking (the crucial difference is the opposite sign in front of $\tan ^{2}$
in the square brackets).

\section{Conclusion}

The objective of this work is to elaborate the simplest setting for the
study of the spontaneous symmetry breaking in dynamical chains. For this
purpose, we have introduced two one-dimensional discrete systems, in the
form of the straight and ring-shaped linear chains, with two symmetrically
inserted nonlinear sites. This pair of sites introduces the symmetry that
may be spontaneously broken. The chains, both rectilinear and circular ones,
can be realized in BEC (with the help of optical lattices) and in optics, in
the form of waveguiding arrays. A full set of analytical solutions for
symmetric, asymmetric, and antisymmetric localized states has been obtained
in the explicit form, for both geometries. The stability of the stationary
modes was investigated through the numerical computation of the eigenvalues
for small perturbations. The SBB\ (symmetry-breaking bifurcation), which is
responsible for the asymmetric modes emerging from the symmetric ones, is of
the subcritical type in the straight lattice, and in the circular one of a
sufficiently large size. The bifurcation becomes supercritical if the system
is made effectively nonlocal, i.e., in the ring of a smaller size, or if the
nonlinearity is spread over relatively broad areas around the two central
sites in the straight chain. The antisymmetric modes are not subject to
bifurcations, although they too may be both stable and unstable. The
development of the instability (when it occurs) was tested with the help of
direct simulations. It was found that the unstable stationary asymmetric
states, which are a part of the subcritical bifurcation, spontaneously
transform into breathers oscillating between the two nonlinear sites, that
may be considered as the restoration of an effective dynamical symmetry.

\end{document}